\documentstyle [12pt]{article}
\makeatletter \oddsidemargin  0in \evensidemargin 0in
\textwidth16cm \RequirePackage[dvips]{graphicx} \textheight 20.5cm
\newcommand{\singlespacing}{\let\CS=\@currsize\renewcommand{\baselinestretch}{1.5}\tiny\CS}
\makeatletter \oddsidemargin  -.1in \evensidemargin -.1in
\newcommand{\doublespacing}{\let\CS=\@currsize\renewcommand{\baselinestretch}{1.35}\tiny\CS}

\def\@citex[#1]#2{\if@filesw\immediate\write\@auxout{\string\citation{#2}}\fi
  \def\@citea{}\@cite{\@for\@citeb:=#2\do
    {\@citea\def\@citea{,\linebreak[0]\hskip0pt plus .2em}%
      \@ifundefined{b@\@citeb}%
    {{\bf ?}\@warning{Citation `\@citeb' on page \thepage\space undefined}}%
      \hbox{\csname b@\@citeb\endcsname}}}{#1}}

\newtheorem{rule-def}[theorem]{Rule}
\begin{document}
\title{\bf Secret Broadcasting of W-type state}\author{I.Chakrabarty $^{1,2}$\thanks{Corresponding author:
E-Mail-indranilc@indiainfo.com },B.S.Choudhury $^2$ \\
$^1$ Heritage Institute of Technology,Kolkata-107,West Bengal,India\\
$^2$ Bengal Engineering and Science University, Howrah, West
Bengal, India }
\date{}
\maketitle{}
\begin{abstract}
In this work we describe a protocol by which one can secretly
broadcast W-type state among three distant partners. This work is
interesting in the sense that we introduce a new kind of local
cloning operation to generate two W- type states between these
partners from a W-type state initially shared by them.
\end{abstract}
\section{Introduction:}
The no-cloning theorem [1], as modified in [4], states that there
is no method to blindly copy a pair of non orthogonal pure states.
More importantly, for any pair of non orthogonal pure states
$\rho_i,i\in \{1,2\}$, there is no trace-preserving completely
positive map $\epsilon$ such that
$\varepsilon(\rho_i)=\rho_i\otimes \rho_i \forall i$. Although
nature prevents us from amplifying an unknown quantum state but
nevertheless one can construct a quantum cloning machine that
duplicates an unknown quantum state
with a fidelity less than unity [1,2,3,4,5,6].\\
Beyond the no-cloning theorem, one can clone an arbitrary quantum
state with some non zero probability [7]. In the past years, much
progress has been made in designing quantum cloning machine.
Buzek-Hillery took the first step towards the construction of
approximate quantum cloning machine [2]. They showed that the
quality of the copies produced by their machine remain same for
all input state. This machine is known as universal quantum
cloning machine (UQCM). Later D. Bruss \textit{et.al} showed this
universal quantum cloning machine to be optimal [5]. After that
different sets of quantum cloning machines like the set of
universal quantum cloning machines, state dependent quantum
cloning machines (i.e. the quality of the copies depend on the
input state) and the probabilistic quantum cloning machines were
proposed.\\
 Entanglement [8], the heart of quantum information
theory, plays a crucial role in computational and communicational
purposes. Therefore, as a valuable resource in quantum information
processing, quantum entanglement has been widely used in quantum
cryptography [9,10],quantum super dense coding [11] and quantum
teleportation [12]. An astonishing feature of quantum information
processing is that information can be "encoded" in non-local
correlations between two separated particles. A lot of work have
been done to extract pure quantum entanglement from partially
entangled state [10]. Now at this point one can ask an question :
whether the opposite is true or not i.e. can quantum correlations
be "decompressed"? The probable answer to this question is
"Broadcasting of quantum entanglement". Broadcasting is nothing
but local copying of non-local quantum correlations. That is the
entanglement originally shared by a single pair is transferred
into two less
entangled pairs using only local operations.\\
Suppose two distant parties A and B share two qubit-entangled
state
\begin{eqnarray}
|\psi\rangle_{AB}=\alpha|00\rangle_{AB}+\beta|11\rangle_{AB}
\end{eqnarray}

Let us assume that the  first qubit belongs to A and the second
qubit belongs to B. Each of these two parties A and B now perform
local cloning operation on their own qubit. It turns out that for
some
values of $\alpha$\\
 (1) non-local output states are
inseparable, and \\
 (2) local output states are separable.\\
V.Buzek et.al. [25] were the first who proved that the
decompression of initial quantum entanglement is possible, i.e.
from a pair of entangled particles, two less entangled pairs can
be obtained by local operations. That means inseparability of
quantum states can be partially broadcasted (cloned) with the help
of local operations. They used optimal universal quantum cloners
for local copying of the subsystems and showed that the non-local
outputs are inseparable if $\alpha^2$ lies in the interval
$(\frac{1}{2}-\frac{\sqrt{39}}{16},\frac{1}{2}+\frac{\sqrt{39}}{16})$.\\
Further S.Bandyopadhyay et.al. [13]  showed that only those
universal quantum cloners whose fidelity is greater than
$\frac{1}{2}(1 + \sqrt{\frac{1}{3}} )$ are suitable because then
the non-local output states become inseparable for some values of
the input parameter $\alpha$. They proved that an entanglement can
be optimally broadcasted only when optimal quantum cloners are
used for local copying and also showed that broadcasting of
entanglement into more than two entangled pairs is not possible
using only local operations. I.Ghiu investigated the broadcasting
of entanglement by using local $1\rightarrow 2$ optimal universal
asymmetric Pauli machines and showed that the inseparability is
optimally broadcast when symmetric cloners are
applied [14].\\
Few years back we studied broadcasting of entanglement using state
dependent quantum cloning machine as a local copier. We showed
that the length of the interval for probability-amplitude-squared
($\alpha^2$) for broadcasting of entanglement using state
dependent cloner can be made larger than the length of the
interval for probability-amplitude-squared for broadcasting
entanglement using state independent cloner [15]. In that work we
showed that there exists local state dependent cloner which gives
better quality copy (in terms of average fidelity) of an entangled
pair than the local universal cloner [15]. In recent past Adhikari
\textit{et.al} in their paper [16] showed that secretly
broadcasting of three-qubit entangled state between two distant
partners with universal quantum cloning machine is possible. They
generalized the result to generate secret entanglement among three
parties. Recently Adhikari et.al proposed a scheme for
broadcasting of continuous variable entanglement [17]. In another
work [18] we presented a
protocol by which one can broadcast five qubit entangled state between three different parties.\\
Along with Einstein-Podolsky-Rosen (EPR)state and
Greenberger-Horne-Zeilinger (GHZ) state, there exist other
entangled states such as W-class states and zero sum amplitude
(ZSA) states [19] which have substantial importance in quantum
information theory. \\
In this work we introduce a new cloning transformation. Each of
 three friends Alice, Bob and Carol is supplied with this cloning
machine so that they can approximately clone their respective
qubits. We start with a W type state of the form
\begin{eqnarray}
|X\rangle_{123}=\alpha|001\rangle_{123}+\beta|010\rangle_{123}+\gamma|100\rangle_{123}
\end{eqnarray}
shared by three distant parties Alice,Bob and Carol. Then each
party apply local approximate cloning machine on their respective
qubits. After that they perform measurements on their respective
machine vectors. Not only that, each party informs others about
their measurement results using Goldenberg and Vaidman's quantum
cryptographic scheme [20] based on orthogonal state. Since the
measurement results are interchanged secretly among them, so Alice
,Bob and Carol share secretly six qubit state. Among six qubit
state, we interestingly find that there exists two three qubit
W-type states
shared by Alice, Bob and Carol.\\
The advantage of this protocol from the previous broadcasting
protocols is that here we secretly generate two states : (1) One
between Alice's original qubit and cloned qubits of Bob and Carol,
(2) Another between original qubits of Bob and Carol with the
cloned qubit of Alice, independent of the input parameters
$\alpha,\beta,\gamma$. Now to have a knowledge about the quantum
information, evesdroppers have to do two things: First, they have
to gather knowledge about the initially shared entangled state and
secondly, they have to collect information about the measurement
result performed by three distant partners. Therefore, the quantum
channel generated by our protocol is more secured and hence can be
used in various
protocols viz. quantum key distribution protocols [23,24].\\\\
\section{Secretly Broadcasting W-type state among three different partners }
In this section we describe our whole protocol below
step by step.\\\\
{\bf Step1: A new Cloning Transformation:}\\\\
 First of all we
introduce a new cloning operation of the form
\begin{eqnarray}
|0\rangle\longrightarrow\frac{1}{\sqrt{x^2+y^2}}(x|00\rangle|\uparrow\rangle+y|10\rangle|\downarrow\rangle)\nonumber\\
|1\rangle\longrightarrow\frac{1}{\sqrt{x^2+y^2}}(x|11\rangle|\uparrow\rangle+y|01\rangle|\downarrow\rangle)
\end{eqnarray}
where $\{|\uparrow\rangle,|\downarrow\rangle\}$ are post operation
orthogonal quantum cloning machine state vectors. Without loss of
generality, $x$ and $y$ can always be considered to be real
parameters. Now each of the three parties are supplied with
identical cloning machines (defined by equation(3)), so that they
can approximately clone their respective
qubits. \\\\
{\bf Step 2: Local Cloning and Measurement}\\\\ Let us consider a
scenario, where three friends Alice, Bob and Carol, who are far
away from each other, are sharing an entangled state (W-type) of
the form
\begin{eqnarray}
|X\rangle_{123}=\alpha|001\rangle_{123}+\beta|010\rangle_{123}+\gamma|100\rangle_{123}
\end{eqnarray}
where $\alpha,\beta,\gamma$ are all real with
$\alpha^2+\beta^2+\gamma^2=1$. The qubits 1,2,3 are with Alice,Bob
and Carol respectively.
\\

Alice , Bob and Carol then operate  quantum cloning machine
defined in equation (3) locally to copy the state of their
respective particles. Therefore, after operating quantum cloning
machine, Alice , Bob and Carol are able to approximately clone the
state of the particle and consequently the combined system of six
qubits is given by
\begin{eqnarray}
&&|X^C\rangle_{142536}={}\nonumber\\&&\frac{1}{(x^2+y^2)^{\frac{3}{2}}}\{\alpha(x|00\rangle|\uparrow\rangle^A+y|10\rangle|\downarrow\rangle^A)(x|00\rangle|\uparrow\rangle^B+y|10\rangle|\downarrow\rangle^B)(x|11\rangle|\uparrow\rangle^C+y|01\rangle|\downarrow\rangle^C){}\nonumber\\&&
+\beta(x|00\rangle|\uparrow\rangle^A+y|10\rangle|\downarrow\rangle^A)(x|11\rangle|\uparrow\rangle^B+y|01\rangle|\downarrow\rangle^B)(x|00\rangle|\uparrow\rangle^C+y|10\rangle|\downarrow\rangle^C){}\nonumber\\&&
+\gamma(x|11\rangle|\uparrow\rangle^A+y|01\rangle|\downarrow\rangle^A)(x|00\rangle|\uparrow\rangle^B+y|10\rangle|\downarrow\rangle^B)(x|00\rangle|\uparrow\rangle^C+y|10\rangle|\downarrow\rangle^C)\}
\end{eqnarray}
The subscripts 4,5,6  refer approximate copies of qubits 1,2,3
which are with Alice, Bob and Carol respectively. Also
$|\rangle^A$ , $|\rangle^B$ and $|\rangle^C$ denotes quantum
cloning machine state vectors in
Alice's , Bob's  and Carol's side respectively\\

Now after local cloning, each of them perform measurement on the
quantum cloning machine state vectors in the basis
$\{|\uparrow\rangle,|\downarrow\rangle\}$ and exchange their
measurement results with each other using Goldenberg and Vaidman's
quantum cryptographic scheme [20] . In this way Alice , Bob and
Carol interchange their measurement
results secretly.\\
 The tensor product of
machine state vectors of three friends after the
measurement is given by the following table.\\\\\\

{\bf TABLE 1:}\\
\begin{tabular}{|c|c|}
\hline Serial Number & Measurement Results   \\
\hline 1 &
$|\uparrow\rangle^A|\uparrow\rangle^B|\uparrow\rangle^C$\\
\hline 2 & $|\uparrow\rangle^A|\uparrow\rangle^B|\downarrow\rangle^C$ \\
\hline 3 & $|\uparrow\rangle^A|\downarrow\rangle^B|\downarrow\rangle^C$ \\
\hline 4
&$|\uparrow\rangle^A|\downarrow\rangle^B|\uparrow\rangle^C$
\\ \hline 5 & $|\downarrow\rangle^A|\uparrow\rangle^B|\uparrow\rangle^C$ \\
\hline 6 & $|\downarrow\rangle^A|\uparrow\rangle^B|\downarrow\rangle^C$\\
\hline 7 & $|\downarrow\rangle^A|\downarrow\rangle^B|\uparrow\rangle^C$\\
\hline 8 & $|\downarrow\rangle^A|\downarrow\rangle^B|\downarrow\rangle^C$\\
\hline
\end{tabular}\\\\
{\bf  Step 3: Analysis of a Particular Measurement Result}\\\\
 Now let us consider the case when the measurement outcome is
$|\uparrow\rangle^A|\uparrow\rangle^B|\uparrow\rangle^C$, then the
six qubit entangled state shared by Alice , Bob and Carol is given
by
\begin{eqnarray}
|Y^C\rangle_{142536}=\frac{x^3}{(x^2+y^2)^{\frac{3}{2}}}\{\alpha|000011\rangle_{142536}+\beta|001100\rangle_{142536}+\gamma|110000\rangle_{142536}\}
\end{eqnarray}
Now it remains to be seen whether one can generate two 3-qubit
W-type state from above six qubit entangled state or not.

\begin{eqnarray}
\rho_{156}=\rho_{234}=\frac{x^4y^2}{(x^2+y^2)^3}\{\alpha^2|001\rangle\langle
001 |+\beta^2|010\rangle\langle 010|+\gamma^2|100\rangle\langle
100 |\nonumber\\+\alpha\beta|001\rangle\langle
010|+\alpha\gamma|001\langle 100 |+\beta\alpha|010\rangle\langle
001|\nonumber\\+\beta\gamma|010\rangle\langle
100|+\gamma\alpha|\rangle100\langle
001|+\gamma\beta|100\rangle\langle 010|\}
\end{eqnarray}
It is evident from the outer products of equation(7), that the
density operators $\rho_{156}$ and $\rho_{234}$ represent the
density
matrix of W-type of states.\\
One can investigate the problem of inseparability of the states
obtained as a consequence of other possible measurement results as
shown in the table 1.
\\\\
{\bf Step 4: Inseparability of Local Output
states}\\\\
In broadcasting of inseparability, we generally use
Peres-Horodecki criteria [21,22] to show the inseparability of
non-local
outputs and separability of local outputs.\\\\

\textbf{Peres-Horodecki Theorem :}The necessary and sufficient
condition for the state $\rho$ of two spins $\frac{1}{2}$ to be
inseparable is that at least one of the eigen values of the
partially transposed operator defined as
$\rho^{T}_{m\mu,n\nu}=\rho_{m\mu,n\nu}$, is negative. This is
equivalent to the condition that at least one of the two
determinants\\
$W_{3}= \begin{array}{|ccc|}
  \rho_{00,00} & \rho_{01,00} & \rho_{00,10} \\
  \rho_{00,01} & \rho_{01,01} & \rho_{00,11} \\
  \rho_{10,00} & \rho_{11,00} & \rho_{10,10}
\end{array}$ and $W_{4}=\begin{array}{|cccc|}
   \rho_{00,00} & \rho_{01,00} & \rho_{00,10} & \rho_{01,10}\\
  \rho_{00,01} & \rho_{01,01} & \rho_{00,11} & \rho_{01,11} \\
  \rho_{10,00} & \rho_{11,00} & \rho_{10,10} & \rho_{11,10} \\
  \rho_{10,01} & \rho_{11,01} & \rho_{10,11} & \rho_{11,11}
\end{array}$\\
is negative.\\\\
Now we have to check that whether in our protocol  the local
output states are separable or not. The density operators
representing the local output states are given by,
\begin{eqnarray}
\rho_{14}=\frac{x^6}{(x^2+y^2)^3}\{\alpha^2|00\rangle\langle 00
|+\beta^2|00\rangle\langle
00|+\gamma^2|11\rangle\langle|11\}\nonumber\\
\rho_{25}=\frac{x^6}{(x^2+y^2)^3}\{\alpha^2|00\rangle\langle 00
|+\beta^2|11\rangle\langle
11|+\gamma^2|00\rangle\langle|00\}\nonumber\\
\rho_{36}=\frac{x^6}{(x^2+y^2)^3}\{\alpha^2|11\rangle\langle 11
|+\beta^2|00\rangle\langle 00|+\gamma^2|00\rangle\langle|00\}
\end{eqnarray}
Now if one applies the Peres-Horodecki criterion to see whether
the states are entangled or not, he will find that for each of
these density operators, $W_4=W_3=0$ independent of values of
$\alpha,\beta,\gamma$. This clearly indicates the fact that
the local output states are separable.\\
Thus with the help of the above protocol one can generate two
three qubit W-type states from
a W-type state:\\
\textbf{(1) \textit{One between Alice's original qubit and cloned qubits of Bob
and Carol.}\\
(2) \textit{Another between original qubits of Bob and Carol with
the cloned qubit of Alice.}}\\
One can use these two secretly broadcasted three qubit W-states
as secret quantum channels between three partners for various
cryptographic schemes.
\section{Conclusion:}In this work, we present a protocol for the secret broadcasting of three-qubit entangled
state (W-type) between three distant partners. Here we should note
an important fact that the two copies of three-qubit entangled
state is  generated from previously shared three-qubit entangled
state independent of the input parameters $\alpha,\beta,\gamma$.
They send their measurement result secretly using cryptographic
scheme so that the produced copies of the three-qubit entangled
state shared between three distant parties can serve as a secret
quantum channel. Another important thing is that instead of
applying (B-H) cloning machine for twice, as in reference [16]
here three parties applied a different cloning transformation .
Now these three parties can use these newly broadcasted W-type
states as quantum channels more securely than any three qubit
entangled states.
\section{Acknowledgement}
I.C acknowledges almighty God for being the source of inspiration
of all work. He also acknowledges Prof C.G.Chakraborti for being
the source of inspiration in research work. I.C also N.Ganguly for
having useful discussions.
\section{References}
$[1]$ W.K.Wootters, W.H.Zurek, Nature 299 (1982) 802.\\ $[2]$
V.Buzek, M.Hillery, Phys.Rev.A 54 (1996) 1844.\\ $[3]$ N.Gisin,
S.Massar, Phys.Rev.Lett. 79 (1997) 2153.\\ $[4]$ H. P. Yuen, Phys. Lett. A 113, 405 (1986).\\
$[5]$ D.Bruss, D.P.DiVincenzo, A.Ekert, C.A.Fuchs, C.Macchiavello,
J.A.Smolin,Phys.Rev.A 57 (1998) 2368.\\
$[6]$ V.Buzek, S.L.Braunstein, M.Hillery, D.Bruss, Phys.Rev.A 56 (1997) 3446.\\
$[7]$ L.M.Duan and G.C.Guo,Phys.Rev.Lett. 80 (1998)4999.\\
$[8]$ Einstein, Podolsky and Rosen, Phys.Rev. 47 (1935)777.\\
$[9]$ C.H.Bennett and G.Brassard, Proceedings of IEEE
International Conference on Computers, System and Signal
Processing, Bangalore, India, 1984, pp.175-179.\\
$[10]$ P.W.Shor and J.Preskill, Phys.Rev.Lett. 85 (2000)441.\\
$[11]$ C.H.Bennett and S.J.Weisner, Phys.Rev.Lett.69 (1992)2881.\\
$[12]$ C.H.Bennett, G.Brassard, C.Crepeau, R.Jozsa, A.Peres
and W.K.Wootters,Phys.Rev.Lett. 70 (1993)1895.\\
$[13]$ S.Bandyopadhyay, G.Kar, Phys.Rev.A 60 (1999)3296.\\
$[14]$ I.Ghiu, Phys.Rev.A 67 (2003)012323.\\
$[15]$ S.Adhikari, B.S.Choudhury and I.Chakrabarty, J. Phys. A: Math. Gen. 39 No 26 (2006)8439.\\
$[16]$ S.Adhikari, B.S.Choudhury, Phys. Rev. A74,  (2006)032323.\\
$[17]$   Satyabrata Adhikari, A. S. Majumdar, N. Nayak
, arXiv:0708.1869.\\
$[18]$ I.Chakrabarty (in preparation).\\
$[19]$ P.W.Shor and J.Preskill, Phys.Rev.Lett. 85 (2000)441.\\
$[20]$ L.Goldenberg and L.Vaidman, Phys.Rev.Lett. 75 (1995)1239.\\
$[21]$ A.Peres, Phys.Rev.Lett. 77 (1996)1413.\\
$[22]$ M.Horodecki, P.Horodecki, R.Horodecki, Phys.Lett.A 223.
(1996)1.\\
$[23]$ A.Cabello, Phys.Rev.A 61 (2000)052312.\\
$[24]$ C.Li, H-S Song and L.Zhou, Journal of Optics B: Quantum
semiclass. opt. 5 (2003)155.\\
$[25]$V.Buzek, V.Vedral, M.B.Plenio, P.L.Knight, M.Hillery,
Phys.Rev.A 55 (1997) 3327.

\end{document}